\documentclass[showpacs,showkeys,amsmath,amssymb,preprint,nofootinbib]{revtex4}
\usepackage{graphicx}
\usepackage{epsfig}
\usepackage{bm}
\usepackage{float}
\usepackage{fontenc}
\usepackage{graphicx}
\usepackage{bm}
\usepackage{amsmath}
\usepackage{xcolor}
\def\beq{\begin{eqnarray}}

\def\eeq{\end{eqnarray}}

\begin{document}

\title[tcc]{Non-zero torsion and late cosmology}

\author{Miguel Cruz$^{a,}$}
\email{miguelcruz02@uv.mx}

\author{Fernando Izaurieta$^{b,}$}
\email{fizaurie@udec.cl}

\author{Samuel Lepe$^{c,}$}
\email{samuel.lepe@pucv.cl}

\affiliation{$^a$Facultad de F\'\i sica, Universidad Veracruzana 91000, Xalapa, Veracruz, M\'exico\\
$^b$Departamento de F\'\i sica, Universidad de Concepci\'on, Casilla 160-C, 4070105, Concepci\'on, Chile\\
$^c$Instituto de F\'\i sica, Facultad de Ciencias, Pontificia Universidad Cat\'olica de Valpara\'\i so, Av. Brasil 2950, Valpara\'\i so, Chile}

\date{\today}

\begin{abstract}
In this work, we study some thermodynamical aspects associated with torsion in a flat FLRW spacetime cosmic evolution. By implementing two Ansatze for the torsion term, we find that the model admits a phantom regime or a quintessence behavior. This scheme differs from the $\Lambda$CDM model at the thermodynamical level. The resulting cosmic expansion is not adiabatic, the fulfillment of the second law of thermodynamics requires a positive torsion term, and the temperature of the cosmic fluid is always positive. The entropy of the torsion phantom scenario is negative, but introducing chemical potential solves this issue. For a Dirac-Milne type Universe, the torsion leads to a growing behavior for the temperature of the fluid but has no incidence on the rate of expansion.       
\end{abstract}


\maketitle
\section{Introduction}
\label{sec:intro}
Our current understanding of the Universe at large scales lies mainly in Einstein's theory of General Relativity. However, the golden age we are living in astrophysical data acquisition unveiled several aspects of the Universe that have forced us to question whether this formulation for gravity is the definitive one or requires extra assumptions. For instance, we still have to understand the nature of the component responsible for the current accelerated cosmic expansion \cite{riess}. A possible solution is to modify the spacetime geometry, as in the case of $f(R)$ gravity \cite{sotiriou}, in this approach an extra degree of freedom given by a scalar field appears providing the possibility of having a consistent unified scenario for the description of early and late times of the Universe \cite{odint1}. This smooth transition from early to late times evolution can be also obtained if an arbitrary function of the torsion scalar, $T$, is considered as gravitational action. An interesting review on this topic can be found in \cite{odint2}. Therefore the identification of torsion as catalyst of cosmic accelerated expansion have been explored exhaustively. As discussed in Ref. \cite{minkevich}, the $\Lambda$CDM model dynamics can be imitated by considering torsion effects in a homogeneous and isotropic spacetime, thus the accelerated expansion of the Universe could be explained only by geometrical considerations of spacetime without invoking the dark sector. The restriction of parameters for a torsional model with the type Ia Supernovae, CMB, BAO and Hubble parameter measurements data is performed in Refs. \cite{bengocheaa, bengocheab}, these studies revealed that this kind of model is viable to describe the current status of the Universe and also exhibits consistently a radiation and matter era followed by an acceleration stage; besides the corresponding value for the matter density parameter obtained in this model is close to the prediction given by the standard cosmology. 

The Riemannian geometry is fully described by the metric, being the connection given the Christoffel symbol. In the case of Riemann-Cartan geometry, the metric and the connection correspond to independent degrees of freedom. In this case, the metric and the contorsion (contortion) tensor locally describe the geometry.

From an experimental point of view, we have no evidence in favor nor against the existence of torsion. In general, it plays the role of a new dark source of torsionless Riemannian gravity. Some authors have even proposed that dark matter could be torsion in disguise Ref.~ \cite{Tilquin:2011bu}.
Its detection through particle physics experiments seems hard (See Ref.~\cite{Puetzfeld:2014sja} and Chap.~8.4 of Ref.~\cite{SupergravityVanProeyen}) but not impossible in the future.

An interesting proposal for the detection of torsion effects by means of the symmetry Lorentz violation in some experiments can be found in Ref.~\cite{proposal}. It is important to point out that this symmetry violation is present in some of the neutrino experiments, therefore there exists the possibility that torsion could play a fundamental role to understand the nature of the neutrinos; specifically on how neutrinos acquire mass or what kind of mass they have (see for instance the Ref. \cite{oscillation}, where it is studied that torsion effects can also be used to understand the neutrino oscillation mechanism). The Lorentz symmetry is contained in the CPT theorem, thus the torsion could be related to the matter- antimatter asymmetry generated in the early universe \cite{neutrinos}.

From a more theoretical point of view, the relation of torsion with the relativistic description of the movement of a supersymmetric particle with spin \cite{zanelli}, in the construction of topological invariants in higher dimensional spaces \cite{zanelli2} or in the study of the conformal symmetry of gravity theory \cite{conformal}.

In the context of Riemann-Cartan geometry, there are two kinds of theories. The closer ones to General Relativity are Einstein-Cartan-Sciama-Kibble (ECSK) theories (Refs~\cite{EinsteinCartanLetters,Kib61,Sciama:1964wt,Hehl:1971qi,doi:10.1142/6742,doi:10.1142/0356,RevModPhys.48.393,Shapiro:2001rz,Hammond:2002rm,Poplawski:2009fb,1793906}.). In this case, the source of torsion is the spin tensor of matter $\sigma^{\lambda}{}_{\mu \nu}$, similarly as the stress-energy tensor $\tau_{\mu \nu}$ is the source of curvature,
\begin{align}
& R_{\mu \nu}-\frac{1}{2}g_{\mu \nu}R+\Lambda g_{\mu \nu}   =\frac{8\pi G}{c^{4}}\tau_{\mu \nu}\,,\label{Eq_metric}\\
& T^{\lambda}{}_{\mu \nu}-\delta_{\mu}^{\lambda}T^{\gamma}{}_{\gamma \nu}+\delta_{\nu}^{\lambda}T^{\gamma}{}_{\gamma \mu}  =\frac{8\pi G}{c^{4}}\sigma^{\lambda}{}_{\mu \nu}\,.\label{Eq_affine}
\end{align}
Since torsion $T^{\lambda}{}_{\mu \nu}$ depends algebraically on $\sigma^{\lambda}{}_{\mu \nu}$, it cannot propagate in a vacuum, and it depends on the energy density through the spin density.

A different option is to look for a different Lagrangian and to depart from GR more dramatically. Some theories in this family are Poincaré Gauge Theory (see Refs.\cite{Hehl1980,Blagojevic:2013xpa}) and nonminimal couplings with topological invariants (see Refs.~\cite{Alexander:2019wne,Magueijo:2019vmk,Barker:2020gcp,Valdivia:2017sat,Cid:2017wtf,Barrientos:2019awg,Alexander:2020umk,Mercuri:2009zi,Obukhov:1997pz,Kreimer:1999yp,Lattanzi:2009mg,Castillo-Felisola:2015ema,Karananas:2018nrj,Castillo-Felisola:2016kpe}). In these kinds of theories, torsion propagates in a vacuum\footnote{For more information on the wave operator on spaces with Riemann-Cartan geometry and the propagation of perturbations, see
Refs.\cite{Barrientos:2019msu,Barrientos:2019awg,Izaurieta:2019dix,Hehl:1979xk,Boos:2016cey}.}, and it is not necessary a spin tensor to have a nonvanishing torsion.
In the current work, we restrict our attention to the first kind of ECSK theories with non-propagating torsion (but closer to GR).

Our aim in this work is to discuss some aspects of the cosmology that results from the torsional formalism when it is implemented for a Friedmann-Lemaitre-Robertson-Walker (FLRW) type spacetime, as in Refs. \cite{kranas, pereira}, and to study a possible relationship between torsion and dark energy.

We found that the corresponding parameter state for the cosmic fluid takes values within the region of quintessence or phantom, this depends on the election of the Ansatz to describe the torsional term. On the other hand, in the thermodynamics description of torsion cosmology we obtain that the cosmic evolution is governed by a non adiabatic behavior for the entropy, this interesting behavior for the entropy is also present in the dark energy - dark matter interaction schemes as well as in cosmological scenarios that consider matter creation \cite{grandon, cardenas}. This is a clear indication that the model is beyond the $\Lambda$CDM model, where the entropy takes a constant value. As we will see later, the review of some thermodynamic aspects in the presence of torsion show that the second law of thermodynamics can be guaranteed for a positive torsion term and the temperature of the fluid is positive. In general grounds we can say that the consideration of torsional effects in the cosmological description does not lead to contradictions with the standard formulation of thermodynamics theory.

This work is organized as follows: In Section \ref{sec:torsion} we discuss the torsion dynamics in a flat FLRW spacetime with the use of a barotropic fluid, as we will see later, the inclusion of torsion in the fluid description can lead to a phantom cosmology under certain conditions. We also discuss two Ansatze for the torsion term found in the literature; while one of them leads to a phantom regime the other only provides a quintessence behavior. The thermodynamics of torsion cosmology is discussed in Section \ref{sec:thermo}. This cosmological model is in agreement with the second law of thermodynamics. The corresponding temperature of the fluid is always positive. However, for a dark matter type fluid and a barotropic temperature the fluid behaves as in the $\Lambda$CDM model, i.e., its temperature remains constant. We consider the inclusion of chemical potential in order to solve the positivity problem of entropy. In this section we also discuss a generalized form for the internal energy known as Komar energy, under this description we found explicit expressions for the pressure and density of the cosmological fluid, in such case a phantom cosmology is allowed and the early Universe does not obey the second law of thermodynamics. In Section \ref{sec:milne} we provide a brief discussion of the Dirac-Milne Universe when torsion is included. In Section \ref{sec:final} we give the final comments of our work.  We will use $8\pi G=c=k_{B}=1$ units throughout this work.

\section{Torsion dynamics}
\label{sec:torsion}
The parity, and the homogeneity and isotropy of a FLRW type spacetime with non-zero torsion are preserved for a torsion tensor of the form \cite{kranas}
\begin{equation}
    S_{abc} = 2\phi h_{a[b}u_{c]},
    \label{eq:torsion}
\end{equation}
where $\phi := \phi(t)$ is a scalar function that depends only on time. Therefore, by considering the standard form of energy-momentum tensor for the matter content and the line element for a flat FLRW Universe; the Friedmann and acceleration equations can be written as follows
\begin{align}
& 3H^{2} = \rho - 12\phi(\phi + H),\label{eq:friedmann}\\
& \dot{H} + H^{2} = -\left[\frac{1}{6}(\rho + 3p)+2(\dot{\phi}+\phi H) \right],\label{eq:accel}
\end{align}
where the quantities $p$ and $\rho$ characterize the pressure and density of the fluid, respectively. Besides, the dot stands for derivative with respect to time and $H$ is the well-known Hubble parameter, which is written in terms of the scale factor as, $H:=\dot{a}/a$. The conservation equation for the energy density of the fluid takes the following form
\begin{equation}
    \dot{\rho} + 3H(\rho + p) + 2\phi(\rho + 3p) = 0,
    \label{eq:cont}
\end{equation}
note that in the absence of torsion we recover the standard cosmology. Using the equations (\ref{eq:friedmann}) and (\ref{eq:accel}) we can construct the deceleration parameter straightforwardly
\begin{equation}
    q = -1 -\frac{\dot{H}}{H^{2}} = \frac{1}{2}\left\lbrace \frac{(1+3\omega)\rho + 12 (\dot{\phi}+\phi H)}{\rho - 12\phi(\phi + H)}\right\rbrace. \label{eq:decel} 
\end{equation}
In general, the pressure and density of the fluid are related as, $p = p(\rho)$; in the previous expression we have considered a barotropic equation of state, i.e., $p = \omega \rho$, which is the most simple assumption. We will refer as parameter state for the constant $\omega$. If the following condition is fulfilled by the parameter state
\begin{equation}
    \omega \leq -1 -\frac{4}{\rho}\left[\dot{\phi}-\phi(2\phi+H) \right],
\end{equation}
therefore the deceleration parameter given in Eq. (\ref{eq:decel}) will obey the condition, $q \leq -1$. For $q=-1$ we have a cosmological constant type evolution and for $q<-1$ the model has an over-accelerated expansion also known as phantom cosmology. It is worthy to mention that the crossing to the phantom regime is due only to the introduction of torsion in the cosmological description of the Universe. On the other hand, if we use the Eqs. (\ref{eq:friedmann}) and (\ref{eq:cont}) we can obtain for the normalized Hubble parameter, $E(t) := H(t)/H_{0}$, where $H_{0}$ is the Hubble constant
\begin{equation}
    E\left( t\right) =\left( \frac{a_{0}}{a}\right) ^{3\left( 1+\omega \right)
/2}\sqrt{\Omega _{\rho }\left( a_{0}\right) \exp \left[ -6\left( \frac{1}{3}+\omega \right) \int_{t_{0}}^{t}\phi \left( t\right) dt\right] }-2\frac{\phi
\left( t\right) }{H_{0}},
\label{eq:hubble}
\end{equation}
where the subscript zero denotes the value of any cosmological quantity at present time, in the previous expression we introduced the density parameter, $\Omega _{\rho }\left( a_{0}\right)$ and it is defined in the usual form as $\Omega _{\rho }\left( a_{0}\right) := \rho(a_{0})/3H^{2}_{0}$. Commonly it is more convenient to express the normalized Hubble parameter as a function of the redshift, this can be done by employing the following relationship, $1+z = a_{0}a^{-1}$, this expression leads to $dt = -[(1+z)H(z)]^{-1}dz = -[(1+z)H_{0}E(z)]^{-1}dz$, then the Eq. (\ref{eq:hubble}) can be written as
\begin{equation}
    E\left( z\right) =\left(1+z\right) ^{3\left( 1+\omega \right)
/2}\sqrt{\Omega _{\rho }\left(0\right) \exp \left[6\left( \frac{1}{3}+\omega \right) \int_{0}^{z} \frac{\phi(z)}{(1+z)H_{0}E(z)} dz\right] }-2\frac{\phi
\left( z\right) }{H_{0}},
\label{eq:hubblez}
\end{equation}
in terms of the redshift the evaluation of the cosmological quantities at present time is given at $z=0$. Note that once we choose a specific form for the torsion term, $\phi$; the Eqs. (\ref{eq:hubble}) and (\ref{eq:hubblez}) become solvable for the normalized Hubble parameter. In order to have explicit expressions for $H$, we will discuss some Ansatz for the torsion term and some of its cosmological consequences in the following section. 

\subsection{Ansatze for $\phi$}

In this section we will consider some Ansatze for the torsion term. As first choice we consider the Ansatz given in Ref. \cite{kranas}, where the torsion term has the form
\begin{equation}
    \phi = \lambda H,
    \label{eq:ansatz1}
\end{equation}
being $\lambda$ a constant that lies in the interval $[-0.005813, 0.019370]$. The bounds for the constant $\lambda$ were constrained with the use of Big Bang Nucleosynthesis data. If we insert the expression (\ref{eq:ansatz1}) in Eq. (\ref{eq:hubblez}) one gets
\begin{equation}
    E(z) = \frac{\sqrt{\Omega_{\rho}\left(0\right)}}{(1+2\lambda)}(1+z)^{\frac{1}{2}\left[3(1+\omega)+6\lambda\left(\frac{1}{3}+\omega \right) \right]}.
    \label{eq:Hansatz1}
\end{equation}

\begin{figure}[htbp!]
\centering
\includegraphics[width=9.4cm,height=7.1cm]{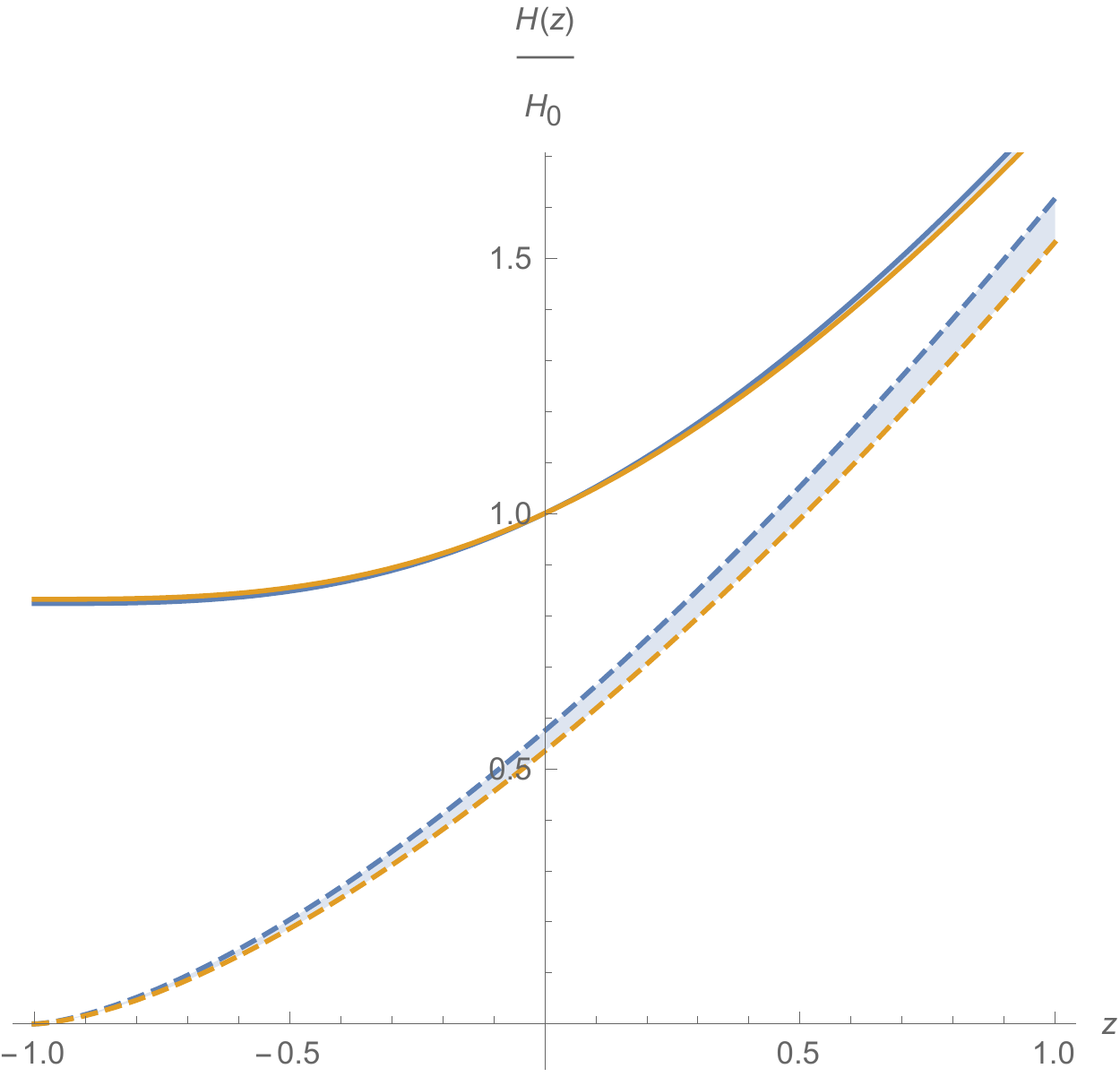}   
\caption{Comparison between $\Lambda$CDM model and non-zero torsion cosmological model.}
\label{fig:comparison}
\end{figure}

In Fig. (\ref{fig:comparison}) we perform a comparison between the normalized Hubble parameter of the $\Lambda$CDM and the one given in Eq. (\ref{eq:Hansatz1}). The region between the dashed lines corresponds to the non-zero torsion cosmological model, notice that the models are {\it closer} at the past ($z>0$). As can be seen in the plot, as we approach to the far future ($z=-1$) the Hubble parameter of the $\Lambda$CDM model tends to a bounded value while for the non-zero torsion model goes to zero, this behavior represents a main difference between the models. In order to compare both models we considered a pressureless fluid (or simply dark matter) given by the condition $\omega = 0$ in Eq. (\ref{eq:Hansatz1}) and the density parameter, $\Omega_{\rho}(0)$, plays the role of the parameter $\Omega_{m,0}$ that appears in the $\Lambda$CDM model. According to the latest Planck results, $\Omega_{m,0} = 0.315 \pm 0.007$ \cite{planck}.\\

If we insert the relationship between the scale factor and the redshift in Eq. (\ref{eq:Hansatz1}), we can obtain an explicit expression for the scale factor as function of time
\begin{equation}
a\left( t\right) =a_{0}\left[ \frac{\left( 1+2\lambda \right)^{2}}{%
\Omega _{\rho }\left( 0\right) H_{0}^{2}}\right] ^{-1/\Delta }\left(t_{s}-t\right) ^{2/\Delta },
\end{equation}
for simplicity in the notation we have defined $\Delta :=3\left(1+\omega\right) +6\lambda \left(1/3+ \omega\right)$ and
\begin{equation}
t_{s}=t_{0}-\frac{2}{\Delta }\sqrt{\frac{\left( 1+2\lambda \right)^{2}}{\Omega _{\rho }\left( 0\right) H_{0}^{2}}}.
\end{equation}
Some comments are in order. For $\Delta < 0$, we have a singular behavior for the scale factor when, $t = t_{s}$; in this case $t_{s}$ represents some time at the future. Note that also the normalized Hubble parameter given in Eq. (\ref{eq:Hansatz1}) diverges as the model evolves towards the future when the condition $\Delta < 0$ is considered; this implies a divergent behavior for the density, $\rho$, and the pressure since we are considering a barotropic equation of state. According to the classification for future singularities provided in Refs. \cite{class1, class2}, these features represent a Big Rip singularity. As discussed before, for the pressureless fluid we have $E(z\rightarrow -1) \rightarrow 0$ and in consequence $\Delta > 0$, therefore no future singularity can be obtained in this case.\\

On the other hand, in Ref. \cite{pereira} an Ansatz for the torsion term is given as a function of the energy density as follows
\begin{equation}
    \frac{\phi(z)}{H(z)} = -\alpha \left(\frac{\rho(z)}{3H^{2}_{0}} \right)^{n},
    \label{eq:ansatz2}
\end{equation}
where $\alpha$ and $n$ are constants that were constrained with the use of Hubble parameter measurements and Pantheon compilation data, yielding $\alpha = 0.14^{+0.14}_{-0.12}$ and $n = -0.47^{+0.23}_{-0.36}$. If we write the continuity equation (\ref{eq:cont}) in its standard form for a barotropic fluid we have
\begin{equation}
    \dot{\rho}+3H\rho(1+\omega_{eff}) = 0,
\end{equation}
where the effective parameter state has the form
\begin{equation}
    \omega_{eff} = \frac{2}{3}\frac{\phi}{H}+\omega\left(1+\frac{2\phi}{H} \right).
\end{equation}
In order to visualize if the Ansatz for the torsion term as the one given in (\ref{eq:ansatz2}) has relevant role in the cosmic evolution we simply compare with the $\Lambda$CDM model, we set $\omega = 0$ in the previous expression and evaluate at present time, one gets
\begin{equation}
    \omega_{eff} = \frac{2}{3}\frac{\phi}{H} = -\alpha \frac{2}{3}\Omega^{n}_{\rho}(0),
    \label{eq:effective}
\end{equation}
where the Eq. (\ref{eq:ansatz2}) was considered. Using the values given before for $\alpha$, $n$ and the density parameter $\Omega_{\rho}(0)$, we find that the effective parameter state lies in the interval, $-0.5 < \omega_{eff} < 0$. Therefore, the Ansatz (\ref{eq:ansatz2}) for the torsion term can not emulate the $\Lambda$CDM model and in some cases the cosmological fluid can behave as quintessence dark energy. From the Friedmann constraint (\ref{eq:friedmann}) we can solve for the $\phi/H$ term, one gets
\begin{equation}
    \frac{\phi}{H} = -\frac{1}{2}\left[1\pm \left(\frac{\rho}{3H^{2}}\right)^{1/2}\right],
    \label{eq:torsion3}
\end{equation}
note that the previous expression is similar to the Ansatz (\ref{eq:ansatz2}), given that we have the condition, $\phi/H < 0$, we will take only the positive branch of the solution. If we insert the obtained $\phi/H$ term (\ref{eq:torsion3}) in Eq. (\ref{eq:effective}) and evaluate at present time, we can write
\begin{equation}
    \omega_{eff} = -\frac{1}{3}\left[1 + \sqrt{\Omega_{\rho}(0)}\right],
\end{equation}
using the values for the density parameter we have for the effective parameter state, $-0.522 < \omega_{eff} < -0.518$, i.e., in this case the cosmological fluid behaves as quintessence dark energy. Therefore from the previous results we have the following statement, an Ansatz of the form as given in (\ref{eq:ansatz2}) or torsion terms proportional to the density of the fluid can not lead to over-accelerated expansion.

\section{Thermodynamics}
\label{sec:thermo}
As starting point we consider the first law of thermodynamics
\begin{equation}
TdS = d(\rho V) + pdV,
\label{eq:first}
\end{equation}
being $V$ the scalar volume\footnote{If we consider a non-zero torsion spacetime its scalar volume takes the form $V = \mathcal{V}+K^{a}{}_{ba}u^{b}$, where $K^{a}{}_{ba}$ is the contortion tensor and $\mathcal{V}$ corresponds to the torsionless counterpart, with a torsion tensor as given in Eq. (\ref{eq:torsion}) we have, $V = 3H + 6\phi = 3H\left(1+\frac{2\phi}{H}\right)$. Note the first term corresponds to the Hubble volume.}, $T$ the temperature of the cosmological fluid and $S$ its entropy. Using the definition of the scalar volume for a non-zero torsion spacetime, we can write the continuity equation (\ref{eq:cont}) as follows
\begin{equation}
    \dot{\rho}+3H\left(1+\frac{2\phi}{H} \right)(\rho+p)-4\phi \rho = 0.
    \label{eq:contvolume}
\end{equation}
If we compute the time derivative of the first law of thermodynamics (\ref{eq:first}), we can write the following expression
\begin{equation}
    \frac{T}{V}\frac{dS}{dt} = 4\phi \rho,
    \label{eq:entro1}
\end{equation}
where we have considered $dV/V = 3H\left(1+\frac{2\phi}{H}\right)dt$, together with the continuity equation (\ref{eq:contvolume}). Note that the cosmic evolution governed by the above expression differs from the standard cosmology at thermodynamics level when torsion it is included, the previous expression indicates that the adiabaticity condition given by, $S = \mbox{constant}$, is no longer available, besides, always that the scalar function $\phi$ remains positive, the entropy will exhibit a positive growth ($dS/dt > 0$) which will be in agreement with the second law of thermodynamics. Alternatively we can compute the change of the entropy via the Gibbs equation
\begin{equation}
    TdS = d\left(\frac{\rho}{n}\right) + pd\left(\frac{1}{n}\right),
\end{equation}
but now we have introduced the density number $n$ \footnote{For a perfect fluid we have, $n^{a}:=nu^{a}$ \cite{maartens}. Therefore the covariant form of the particle conservation is given by $\nabla_{a}n^{a} = - \dot{n}u_{a}u^{a} + nV = 0$, where we have considered $\nabla_{a}u^{b} = (V/3) h_{a}{}^{b}$ as deduced in Ref. \cite{kranas} for a FLRW type spacetime, being $h_{a}{}^{b}$ the projection tensor defined as $h_{ab} := g_{ab}+u_{a}u_{b}$ and $h_{a}{}^{a} =3$, $u_{a}u^{a} = -1$. As commented previously, $V$ is the scalar volume and corresponds to a non-zero torsion spacetime.}. Then we can write
\begin{equation}
    nTdS = -\left(\rho+p\right)\frac{dn}{n} + d\rho,
    \label{eq:varia}
\end{equation}
taking the time derivative of the previous expression one gets
\begin{equation}
    nT\frac{dS}{dt} = -\left(\rho+p\right)\frac{\dot{n}}{n} + \dot{\rho} = 4\phi \rho,
    \label{eq:entro2}
\end{equation}
where we have considered the particle conservation equation and the continuity equation given in (\ref{eq:contvolume}) for the density, this result coincides with Eq. (\ref{eq:entro1}), as expected. From Eqs. (\ref{eq:entro1}) and (\ref{eq:entro2}) we can observe that torsion acts as a source for the entropy production. The adiabatic expansion of the Universe can be recovered in the zero torsion case. Note that the r.h.s. of Eq. (\ref{eq:varia}) suggests the functional form for the temperature, we can consider $T = T(n,\rho)$. We can compute for the temperature
\begin{equation}
    \dot{T} = \frac{\partial T}{\partial n}\dot{n}+\frac{\partial T}{\partial \rho}\dot{\rho},
\end{equation}
and from the previous expression we have the following temperature evolution\footnote{From the integrability condition
\begin{equation*}
    \frac{\partial^{2}S}{\partial T \partial n} = \frac{\partial^{2}S}{\partial n \partial T},
\end{equation*}
the Gibbs equation (\ref{eq:varia}) becomes \cite{maartens}
\begin{equation*}
    n\frac{\partial T}{\partial n}+(\rho +p)\frac{\partial T}{\partial \rho} = T\frac{\partial p}{\partial \rho}.
\end{equation*}
} 
\begin{equation}
\frac{\dot{T}}{T} = 4\phi \rho \left(\frac{\partial T}{\partial \rho} \right)-3H\left(\frac{\partial p}{\partial \rho}\right) = 4\phi \rho \left(\frac{\partial T}{\partial \rho} \right)-3H\omega \left(1+\frac{2\phi}{H}\right), 
\label{eq:evolution}
\end{equation}
where we have considered a barotropic equation of state. Always that the Gibbs integrability condition holds together with the number ($n$) and energy conservation, the previous expression will be valid. As can be seen, the resulting temperature will depend of the density, $\rho$, then we say we have a barotropic temperature, the most simple assumption for this kind of temperature is $T(\rho) \propto \rho^{\omega/(1+\omega)}$ \cite{maartens}. However, other more general forms for the temperature can appear. Using the relation between the redshift and the scale factor we can write the evolution equation (\ref{eq:evolution}) in terms of the redshift as follows
\begin{equation}
    T(z) = T_{0}\exp \left[ - 4\int \frac{\rho \phi}{H(z)}\left(\frac{\partial T}{\partial \rho} \right)\frac{dz}{(1+z)}+3\omega \int \left(1+\frac{2\phi}{H(z)} \right)\frac{dz}{(1+z)} \right].
\end{equation}
If we consider the Ansatz given in (\ref{eq:ansatz1}) for the torsion term the previous expression can be simplified, yielding
\begin{equation}
    T(z) = T_{0}(1+z)^{\alpha}\exp\left[-4\lambda \int \rho \left(\frac{\partial T}{\partial \rho} \right)\frac{dz}{(1+z)} \right],
    \label{eq:temp}
\end{equation}
where we have defined $\alpha := 3\omega (1+2\lambda)$. It is worthy to mention that for the null torsion case ($\lambda = 0$) the above equation reduces to the standard expression for the temperature, i.e., $T(z) = T_{0}(1+z)^{3\omega}$. Besides, the temperature defined in (\ref{eq:temp}) will be always positive and becomes singular at the far future for $\alpha < 0$. For the dark matter case, $\omega = 0$, we have $\alpha = 0$. However, the temperature is not a constant. This is another difference with the $\Lambda$CDM model, where $T_{dm} = \mbox{constant}$. The variating behavior for the temperature of dark matter seems to be a more consistent description from the thermodynamics point of view, see Refs. \cite{grandon, cardenas}, where this kind of behavior was obtained for the dark matter temperature. Finally, by considering the must simple form for the barotropic temperature in Eq. (\ref{eq:temp}) leads to
\begin{equation}
    T(z) = T_{0}(1+z)^{\alpha}\exp\left[-\frac{4\lambda \omega}{(1+\omega)} \int \rho^{\omega/(1+\omega)} \frac{dz}{(1+z)} \right],
    \label{eq:temp2}
\end{equation}
then we have that $T=T_{0}$ for $\omega = 0$, i.e., the same behavior as in the $\Lambda$CDM model for the dark matter temperature; this means that under the must simple assumption for the barotropic temperature, the dark matter thermodynamics it is not affected by the torsion effects.\\

Given that the temperature of the cosmological fluid discussed previously it is always positive, from the Euler relation  \cite{callen} with a barotropic equation of state one gets
\begin{equation}
    TS = (1+\omega)\rho V,
\end{equation}
and as can be observed for the phantom regime we have $TS <0$ since $\omega <-1$, this implies a negative entropy for the phantom cosmology. Within the scheme of standard cosmology was found that the aforementioned problem for the entropy can be solved if a chemical potential, denoted as $\mu$, it is introduced at cosmological level \cite{termo3, lima1}, i.e., the Euler relation takes the form
\begin{equation}
    TS = (1+\omega)\rho V-\mu N,
\end{equation}
where $N$ is the number of particles contained in the volume $V$, $N = nV$. Always that $\mu > - \left|1+\omega \right|(\rho V)/N$ we will have, $TS > 0$, as expected in standard thermodynamics. Note that in our description we must also include chemical potential in order to avoid the negativity entropy problem for the phantom scenario. With the inclusion of chemical potential the first law (\ref{eq:first}) reads
\begin{equation}
    TdS = d(\rho V) + pdV-\mu dN,
    \label{eq:fchemical}
\end{equation}
and given the usual interpretation for the chemical potential, the conservation of the density number must be modified to $\dot{n}+nV = \nu n$ or in terms of the number $N$ we have $\dot{n}/n+V = \nu = \dot{N}/N$, where $\nu$ is the particle production (annihilation) rate if $\nu > 0 \ (\nu < 0)$. Taking the time derivative of Eq. (\ref{eq:fchemical}) and using the equations (\ref{eq:contvolume}), (\ref{eq:entro2}) together with the modified particle number conservation equation we can write
\begin{equation}
    \mu = \frac{4\phi \rho}{n \nu}\left(\frac{N-1}{N}\right).
\end{equation}
Note that for null torsion the chemical potential vanishes, besides the chemical potential can turn negative in some cases, for annihilation of particles or if $\phi <0$. However, this latter case can be discarded since this condition leads to a negative growth for the entropy (see Eqs. (\ref{eq:entro1}) and (\ref{eq:entro2})). Therefore, if the following condition is satisfied
\begin{equation}
    \frac{4\phi}{\nu}\left(\frac{N-1}{N}\right) > - \left|1+\omega \right|,
\end{equation}
the positivity of the entropy is guaranteed. As commented before, the introduction of chemical potential leads to a well defined thermodynamics for the phantom regime. See for instance the Refs. \cite{chemical1, chemical2} where the introduction of chemical potential in models beyond the standard cosmology resolves the negativity problem of entropy or temperature in a phantom scenario.

\subsection{Generalized form of the energy}
In this section we will briefly discuss a generalization for the energy expression that appears in the first law of thermodynamics. Besides, in order to find some solutions we will consider the Ansatz given in Eq. (\ref{eq:ansatz1}) for the torsion term. Usually can be found that the internal energy is given by the Misner-Sharp term, $U_{MS} = \rho V$, see Eq. (\ref{eq:first}). However, some works show that this expression for the energy can lead to thermodynamics inconsistencies when it is applied to describe an expanding Universe. A simple generalization for the energy that overcomes some of the thermodynamics inconsistencies is the Komar energy, which is given by $U_{K} = (\rho + 3p)V$, see for instance the Refs. \cite{komar1,komar2}, where some cosmological features of this energy are explored in detail. Note that when the Komar energy is considered we are taking into account the effects of the cosmological fluid pressure, therefore we have a more realistic description of the cosmic evolution. We start from the standard definition for the pressure \cite{callen}
\begin{equation}
    p = - \left(\frac{\partial U}{\partial V}\right),
    \label{eq:press}
\end{equation}
and the previous expression is valid always that the number of particles is conserved. On the other hand, from the Friedmann constraint (\ref{eq:friedmann}) we can have a simple expression between the energy density and the scalar volume given as $\rho = V^{2}/3$, therefore if we consider that $\frac{\partial}{\partial V} = \frac{\partial \rho}{\partial V}\frac{\partial}{\partial \rho}$, we can write the following differential equation for the pressure by considering the Komar energy in (\ref{eq:press})
\begin{equation}
    \frac{dp}{d\rho}+\frac{2}{3}\frac{p}{\rho}+\frac{1}{2} = 0,
\label{eq:diff}
\end{equation}
yielding the solution
\begin{equation}
    p(\rho) = \rho \left(\frac{c_{1}}{\rho^{5/3}}-\frac{3}{10} \right),
    \label{eq:press1}
\end{equation}
 where $c_{1}$ is an integration constant and we have assumed that $p = p(\rho)$ in Eq. (\ref{eq:diff}). Now, if we insert the obtained pressure in the continuity equation (\ref{eq:contvolume}), we obtain for the density
 \begin{equation}
     \rho(a) = \rho_{0}a^{-\gamma}\left[\frac{1-3c_{1}(1+2\lambda)a^{\gamma}}{\gamma}\right]^{3/5},
     \label{eq:dens1}
 \end{equation}
where we have defined $\gamma := \frac{21}{10}\left(1+\frac{2\lambda}{21}\right)$, for simplicity in the notation. By means of the previous expression for the density we can compute straightforwardly the Hubble parameter using the Friedmann constraint (\ref{eq:friedmann})
\begin{equation}
    H^{2}(a) = \frac{\rho(a)}{3(1+2\lambda)^{2}}.
\end{equation}
Then, for a barotropic fluid, $p = \omega \rho$, by means of Eqs. (\ref{eq:press1}) and (\ref{eq:dens1}) we can write 
\begin{equation}
    \omega(z) = \frac{p}{\rho} = \frac{c\gamma}{(1+z)^{(2\gamma)/3}\left[(1+z)^{\gamma}-3c_{1}(1+2\lambda)\right]}-\frac{3}{10}, 
\end{equation}
 where the usual relation between the scale factor and the redshift was used and $c := c_{1}/\rho^{5/3}_{0}$ is a constant. It is worthy to mention that as the model evolves to the future ($z \rightarrow -1$), the term $3c_{1}(1+2\lambda)$ dominates over $(1+z)$, therefore the parameter state becomes negative and $\left|\omega\right| > 1$, i.e., this model crosses the phantom divide. As can be seen, the torsion effects have a direct contribution on the parameter state $\omega$.\\ 
 
 We can also compute the behavior of the entropy from the first law of thermodynamics but now considering the Komar energy in the Eq. (\ref{eq:first}), one gets
 \begin{equation}
     \frac{T}{V}\frac{dS}{dt} = 4\phi \rho + 3\left[\dot{p}+3Hp\left(1+\frac{2\phi}{H}\right) \right],
 \end{equation}
 where we used the form of the continuity equation given in (\ref{eq:contvolume}). We can write the previous expression in equivalently as follows
 \begin{equation}
     \frac{T}{V}\frac{dS}{da} = 4\lambda \frac{\rho}{a}+3\left[\frac{dp}{d\rho}\frac{d\rho}{da}+3\frac{p}{a}\left(1+2\lambda \right) \right].
\end{equation}
This model is in agreement with the second law of thermodynamics ($dS/da > 0$) as the scale factor grows, from the results obtained previously for the pressure and density we have a negative growth for the entropy as $a \rightarrow 0$. Therefore when torsion effects are considered, the second law of thermodynamics it is not fulfilled in the early Universe. 

\section{The Dirac-Milne Universe}
\label{sec:milne}
In this section we will discuss some generalities of the Dirac-Milne Universe. This kind of Universe it is characterized by a non accelerated expansion, $\ddot{a} = 0$, or in other words null deceleration parameter. In standard cosmology ($\phi = 0$), the acceleration equation given in the expression (\ref{eq:accel}) takes the following form for a barotropic fluid
\begin{equation}
    \frac{\ddot{a}}{a} = -\frac{1}{6}\left(1 + 3\omega \right)\rho,
\end{equation}
and the deceleration parameter (\ref{eq:decel}) can be written as, $q=(1+3\omega)/2$, then for $\omega = -1/3$ we have $\ddot{a} = q = 0$. On the other hand, taking into account the aforementioned value for the parameter state in the torsion cosmology we have that the normalized Hubble parameter (\ref{eq:Hansatz1}) can be simplified to
\begin{equation}
    E(z) = \frac{\sqrt{\Omega_{\rho}(0)}}{(1+2\lambda)}(1+z).
\label{eq:Emilne}
\end{equation}
By considering the definition of the deceleration parameter
\begin{equation}
    1+q(z) = (1+z)\frac{d \ln E(z)}{dz}, 
\end{equation}
and the Eq. (\ref{eq:Emilne}) one gets $q(z) = 0$. Therefore, the inclusion of torsion in the cosmological description maintains unaltered the Dirac-Milne Universe. In this case the temperature of the fluid given in Eq. (\ref{eq:temp}) has a growing behavior and becomes singular at the far future given that $\alpha < 0$. It is worthy to mention that torsion effects do not have incidence on the acceleration expansion of the Dirac-Milne Universe. This differs from other extended theories where the incorporation of additional effects in the fluid can modify the evolution of the Dirac-Milne type Universe; see for instance the Ref. \cite{norman}, where the inclusion of dissipative effects leads to a deceleration parameter different from zero in the Dirac-Milne Universe.     
 
\section{Final remarks}
\label{sec:final}
At the moment, the most successful theoretical proposal to describe the current state of the Universe is the $\Lambda$CDM model. It depends on two components: cold dark matter ($\omega = 0$), crucial to forming structures, and dark energy ($\omega = -1$) to drive the accelerated cosmic expansion.

According to the standard thermodynamical scheme, these fluids should not interact, and the cosmic evolution should be adiabatic, i.e., the entropy is a constant quantity. However, this picture of the Universe lacks physical consistency. One of the possibilities to alleviate this issue is given by allowing dark matter - dark energy interaction in standard cosmology or going to descriptions beyond the standard cosmological model. 

Based on the results obtained in this work, we observe that torsion in cosmic dynamics behaves similarly to the interacting dark matter - dark energy scheme, leading to non-adiabaticity for the cosmic expansion. Moreover, the fulfillment of the second law of thermodynamics requires a positive torsion term. This feature is essential, given that everything seems to indicate that the second law is still obeyed by the Universe at large scales \cite{manuel}. Even further, the fluid temperature must be positive independently of the parameters of the model. On the other hand, under certain circumstances, the inclusion of torsion can lead to a phantom or quintessence behavior, depending on the Ansatz considered for the torsion term, $\phi/H$. As found in the literature for other cosmological models, the phantom torsion regime has the negativity problem for the entropy, but considering the chemical potential in the thermodynamical picture solves this issue. A generalization of the Misner-Sharp energy, $\rho V$, was also considered, we focused on the Komar energy which is given by $(\rho + 3p)V$, this represents a more realistic scenario at thermodynamics level. Within this context, the parameter state of the fluid crosses the phantom divide and corresponds to a function of the torsion parameter. In this case, the second law of thermodynamics breaks in the early Universe. It is worthy of mentioning that Komar energy is not the unique choice to generalize the internal energy.

Another remarkable result is that the dark matter temperature is not affected by the torsion term when one considers a barotropic temperature, i.e., as in the $\Lambda$CDM model, the dark matter temperature has a positive and constant value. Additionally, a brief review of the Dirac-Milne Universe reveals that, in the presence of torsion, its temperature shows a growing behavior. However, contrary to what happens in other modified gravity theories, the value of the deceleration parameter for this Universe vanishes.

Many features of torsion deserve more research. One of them is its possible role as dark energy and dark matter, or as a component of dark energy and dark matter. For instance, Ref.~\cite{Izaurieta:2020xpk} analyzed a simple model of dark matter with nonvanishing spin tensor. The torsion arising from it behave as a dark matter amplifier and solved the Hubble parameter tension, but much more complex scenarios are also possible.
A possible way to decide these kinds of questions is through gravitational waves. Refs.~\cite{Valdivia:2017sat, Barrientos:2019msu, Barrientos:2019awg} proved that a torsional background changes the propagation of the polarization and amplitude of a gravitational wave, leaving its dispersion relation untouched. It opens the possibility to distinguish torsion from other dark components using gravitational waves as probes.

\section*{Acknowledgments}
M.C. work has been supported by S.N.I. (CONACyT-M\'exico). F.I. acknowledges financial support from the Chilean government through FONDECYT grant 1180681 of the Government of Chile.

\end{document}